\documentstyle[12pt,a4,epsfig,epsf]{article}    
\newcommand{\beq}{\begin{equation}}

\newcommand{\eeq}{\end{equation}}

\newcommand{\beqar}[1]{\begin{eqnarray}\label{#1}}

\newcommand{\eeqar}{\end{eqnarray}}

\relax

\begin{document}    
\title{An Analytical Expression for the Non-Singlet 
Structure Functions \\ at Small $x$ in the Double Logarithmic
 Approximation  \\  }       
\author{ Michael ~Lublinsky \\  
 DESY Theory Group, DESY \\    
 Notkestr. 85, 22607 Hamburg, Germany \\   
E-mail: lublinm@mail.desy.de}    
\maketitle

%\thispagestyle{empty}

%\begin{abstract}
\abstract{
A simple analytic expression for the non-singlet structure function
$f_{NS}$ is given. The expression is derived from the result of Ref. \cite{Ry}
obtained by low $x$ resummation of the quark ladder diagrams 
in the double logarithmic approximation of perturbative QCD.
%\end{abstract}
}

%\keywords{}   
    \begin{flushright}
\vspace{-14.5cm}
DESY 04-002\\
\end{flushright}
\vspace{12.5cm}

\newpage
\setcounter{equation}{1}

The small $x$ behavior of the non-singlet structure functions
in DIS are of crucial importance for the determination of the quark 
densities. As an example, a verification  of the Gottfried sum rules
requires an extrapolation of the experimental data to region of low $x$
\cite{kataev} which can be done using non-singlet structure functions.

The small $x$ behavior of the non-singlet structure functions $f_{NS}$ 
was studied in Ref. \cite{Ry}. Using the method of infrared evolution
equation (IREE)  the pQCD double logarithmic contribution to $f_{NS}$ was
calculated. The result was shown to differ dramatically from the one
predicted by the low $x$ approximation of the conventional DGLAP equation.    
The non-singlet structure function $f_{NS}$ found in Ref. \cite{Ry}
was presented in the form of a complex $\omega$ integral, the
inverse Mellin transform  of a given partial wave. Though one can use this
result to make general estimates and read off the leading asymptotics, it is
not really ``user friendly'' and hard for numerical implementations. 

Recently a detailed study of the quark ladder double logarithmic resummations 
was carried out in Ref. \cite{BL}. Several methods were analyzed and compared.
In particular, a comparison of a direct diagrammatic resummation using 
Bethe-Salpeter equation with IREE leads to a new insight in the structure
of the latter. A particular outcome of this comparison is a clear 
understanding of the way the $\omega$ integrations should be performed. 

In the present paper we apply the results of analysis of Ref. \cite{BL}
to the results of Ref. \cite{Ry} and derive a closed form expression 
for $f_{NS}$. Fortunately, as observed in Ref. \cite{BL}, a unique choice
of a closed $\omega$ integration path allows the integration to be 
performed giving rise to a resulting expression in terms of modified Bessel 
functions $I_\nu$. This expression
is most easy to handle in any further applications.

The structure function $f_{NS}$ is proportional to the imaginary part
of the forward photon scattering amplitude  projected onto the flavor 
non-singlet state. In the double logarithmic approximation of pQCD this
amounts for summing up of diagrams built of two $t$-channel quarks and
$s$-channel gluons forming a ladder. The quark ladders can be resummed
using the IREE in partial wave representation. 
             
We define the  partial wave through the ansatz:   
\beq\label{pw}   
M(Q^2,\,s)\,=\,\int    
\frac{d\omega}{2\,\pi\,i}\,\left(\frac{s}{\mu^2}\right)^\omega\,
\tilde F(\omega,\,Q^2/\mu^2)\,.  
\eeq    
The function $\tilde F(\omega,\,Q^2/\mu^2)$ is a  partial wave representation
of the $s$-channel contribution to the quark-photon elastic  amplitude $M$. 
The imaginary part is obtained from the $u$-crossed amplitude which at high
energies approximately equals $M(Q^2,\,-s)$.
The scale $\mu$  is introduced as an auxiliary infrared cutoff parameter, 
the minimal quark transverse momentum.
For a more detailed description of the    
method of IREE we refer the reader to the original work \cite{KiLi} 
as well as to some applications in Ref. \cite{Ry,BER1,BER2}.

The IREE for $\tilde F$ is obtained by differentiating (\ref{pw}) 
with respect to  $\ln \mu^2$.   
The equation reads    
\beq\label{IREE}   
\omega\,\tilde F\,\,+\,\,\frac{\partial \tilde F}{\partial y}
\,\,=\,\,
\,\,\frac{1}{8\,\pi^2}\,  f_0(\omega)\,\,\tilde F(\omega, y)
\eeq    
with $y\,\equiv\,\ln (Q^2/\tilde\mu^2)$, 
The function $f_0$ was introduced in Ref. \cite{KiLi}     
\beq\label{f0}    
f_0(\omega)\,=\,4\,\pi^2\,(\omega\,-\,\sqrt{\omega^2-\omega_0^2})\,;\,\,\,\,\,
\,\,\,\,\,\,\,\,\,\,\,\,\,\,\,\omega_0^2\,=\,2\,\alpha_s\,C_F/\pi    
\eeq     
The function $\tilde F(\omega,\, y)$ solving the IREE (\ref{IREE})    
has the form \cite{KiLi,Ry}:
\beq\label{F}    
\tilde F(\omega,\,y)\,=\,\frac{C_0}{g^2\,C_F}\,f_0(\omega)\,
e^{\,-(\omega\,-\,f_0/8\pi^2)\,y}    
\,\,\,\,\,\,\,\,\,\,\,\,\,\,\,\,for\,\,\,\,\,\,\,\,y \ge 0\,,  
\eeq    
The coefficient $C_0$    
contains information about coupling of the photon to the ladder, 
$C_0\,=\,-\,4\,\pi\,\alpha_{em}\,\epsilon (A) \,\cdot\,\epsilon (A^\prime)$, 
with $\epsilon$ being the photon polarization vectors. In the 
Born approximation
the above choice of $C_0$ corresponds to a scattering off a free quark
with initial condition proportional to $\delta(1\,-\,x)$.

The $s$-channel part of the elastic amplitude $M$ reads    
\beq\label{pw1}   
M(Q^2,\,s,\,\tilde \mu)\,=\,\frac{C_0}{g^2\,C_F}\,\int    
\frac{d\omega}{2\,\pi\,i}\,\left(\frac{s}{Q^2}\right)^\omega\,
 f_0(\omega)\,exp[\,y\,f_0(\omega)/8\pi^2]\,.  
\eeq    
The non-singlet structure function $f_{NS}$ is related to $M$ via
\beq\label{fns0}
-\,\pi\,\,C_0\,\,f_{NS}\,\,=\,\,e_q^2\,\,Im \,M\,\,\simeq\,\,-\,\pi\,e_q^2\,
\frac{\partial\, M}{\partial\,\ln s}
\eeq
and this results in the following expression
\beq\label{fns}   
f_{NS}\,=\,\frac{e_q^2}{g^2\,C_F}\,\int    
\frac{d\omega}{2\,\pi\,i}\,\left(\frac{s}{Q^2}\right)^\omega\,
\omega\, f_0(\omega)\,exp[\,y\,f_0(\omega)/8\pi^2]\,.  
\eeq   
The results above are essentially copied from Ref. \cite{Ry}. What is our new
observation is that we can proceed further in performing the $\omega$ 
integrations in Eqs. (\ref{pw1}) and (\ref{fns}). The procedure is
very similar to the one described in the Appendix B of Ref. \cite{BL}.

As was emphasized in Ref. \cite{BL} the $\omega$-integration path $C_{cut}$   
goes around the square root branch cut     
from $-\omega_0$ to $\omega_0$.    
\beq\label{T2}    
M\,=\,\frac{C_0}{g^2\,C_F}\,4\,\pi^2\,\int_{C_{cut}}    
\frac{d\omega}{2 \pi i} \left(\frac{s}{    
Q^2}\right)^\omega\,(\omega\,-\,\sqrt{\omega^2-\omega_0^2})\,    
\left(\frac{Q^2}{\mu^2}\right)^{(\omega\,-\,\sqrt{\omega^2-\omega_0^2})/2}    
\eeq    
Denote by $\xi\,=\,\ln (s\,/Q^2)$ and $\eta\,=\,\ln (s/\mu^2)$.    
Let us introduce a new complex variable    
\beq\label{ztrans}    
z\,\,=\,\, (\omega\,-\,\sqrt{\omega^2\,-\,\omega^2_0})/\omega_0\,.    
\eeq    
The $\omega$ integral then turns into a contour integral in the $z$-plane,    
and the integration path encircles the origin $z\,=\,0$.
We obtain:    
\beq\label{T22}    
M\,=\,2\,\pi^2\,\omega_0^2\,\frac{C_0}{g^2\,C_F}\,\left[ \int_{0^+}    
\frac{dz}{2\,\pi\,i}\,\left(\frac{1}{z}\,-\,z\right)\,    
e^{\eta\,\omega_0\,z/2\,+\,\xi\,\omega_0/2\,z}\right] \,=\,    
\eeq    
$$
C_0\,\left(\,    
I_0(\omega_0\,\sqrt{\xi\,\eta})\,-\,\frac{\xi}{\eta}\,
I_2(\omega_0\,\sqrt{\xi\,\eta})\,\right)\,.    
$$
The expression (\ref{T22}) is very similar to the one found
in Ref. \cite{BL} for the non-singlet structure function of photon.

Finally, the expression for $f_{NS}$ is given by
\beq\label{fns1}
f_{NS}\,=\,\frac{e_q^2\,\pi^2\,\omega_0^3}{g^2\,C_F}\,\,\left[ \int_{0^+}    
\frac{dz}{2\,\pi\,i}\,\left(\frac{1}{z^2}\,-\,z^2\right)\,    
e^{\eta\,\omega_0\,z/2\,+\,\xi\,\omega_0/2\,z}\right] \,=\,    
\eeq    
$$
\frac{e_q^2\,\omega_0}{2}\,\left 
(\frac{\sqrt\eta}{\sqrt\xi}\,
I_1(\omega_0\,\sqrt{\xi\,\eta})\,-\,\frac{\xi^{3/2}}{\eta^{3/2}}\,
I_3(\omega_0\,\sqrt{\xi\,\eta})\,\right)\,.
$$   
Eq. (\ref{fns1}) is the main result of this paper. In a somewhat different
context a very similar expression can be also found in Ref. \cite{IKMT}.
The expression (\ref{fns1}) is correct for positive $\xi$ and $\eta$.
At $x\,=\,1$ ($\xi\,=\,0$), 
the $\omega$ integral is divergent, which corresponds
to $\delta(1-x)$ initial condition.

An additional comment is in order. Strictly speaking the result obtained
above should be called a structure function of a free quark rather than 
of a proton. This is because a particular choice of the initial conditions
was implemented. We believe, however, that in the Regge limit considered
above factorization holds implying the  experssion (\ref{fns1}) is a good 
approximation for a proton structure modulo an overall normalization constant.
An alternative approach to the problem based on the NLO DGLAP equation 
and a low scale convolution can be found in Ref. \cite{BV}.  

In the Regge limit $\ln s/Q^2\,=\,\ln 1/x\,\gg \,\ln Q^2/\mu^2\,\gg\,1$ the 
leading asymptotics of the Bessel functions coincide and cancel.  
\beq\label{regge}
f_{NS}\,\sim\,\left(\frac{1}{x}\right)^{\omega_0}
\frac{e_q^2}{\ln^{3/2}(1/x)}
\left(\frac{Q^2}{\mu^2}\right)^{\omega_0/2}
\eeq
Up to the pre-exponential factor 
this asymptotics was correctly found in Ref. \cite{Ry}. 
Exploring another limit
$\ln Q^2/\mu^2\,\gg\,\ln 1/x\,\gg \,1$ we reproduce the low $x$ approximation
of the DGLAP equation 
\beq\label{dglap}
f_{NS}\,\sim\,
\frac{e_q^2}{\ln(1/x)}\,e^{\,\omega_0\sqrt{\ln(1/x)\,\ln Q^2/\mu^2}}
\eeq
Note that only the first term ($I_1$) contrubutes to this asymptotics. 
Eq. (\ref{fns1})  gives an analytic expression for the non-singlet 
structure function $f_{NS}$ and it provides a smooth interpolation 
between two high energy limits. It is worth a comment, however,
 that the low $x$ approximation of the DGLAP equation is reproduced 
with a constant coupling constant only. An extensive work on inclusion
of the running coupling effects has been done in Ref. \cite{EGT}.

%\acknowledgments{
\section*{Acknowledgments}
The author is very grateful to Jochen Bartels for reading of the 
manuscript and
for the fruitful  collaboration in Ref. \cite{BL} which lead to the 
observation reported in this paper. 
\\
The author would like to thank  Andrei Kataev for  a very informative and 
stimulating discussion. 
\\
The valuable comments from 
          Johannes Bluemlein, Boris Ermolaev and  Yura Kovchegov
are greatly acknoledged.
%}  

\end{document}